# Estimation Procedures for Robust Sensor Control

Greg Hager and Max Mintz*
General Robotics and Active Sensory Processing Laboratory
Department of Computer and Information Science
University of Pennsylvania
Philadelphia, PA 19104-6389

## Abstract

Many robotic sensor estimation problems can characterized in terms of nonlinear measurement systems. These systems are contaminated with noise and may be underdetermined from a single observation. In order to get reliable estimation results, the system must choose views which result in an overdetermined system. This is the sensor control problem.

Accurate and reliable sensor control requires an estimation procedure which yields both estimates and measures of its own performance. In the case of nonlinear measurement systems, computationally simple closed-form estimation solutions may not exist. However, approximation techniques provide viable alternatives. In this paper, we evaluate three estimation techniques: the extended Kalman filter, a discrete Bayes approximation, and an iterative Bayes approximation. We present mathematical results and simulation statistics illustrating operating conditions where the extended Kalman filter is inappropriate for sensor control, and discuss issues in the use of the discrete Bayes approximation.

## 1 Introduction

Our ability to build intelligent robots will in large part depend on methods for making decisions and planning relative to a sensor-based world model. This decision-making is complicated by the unavoidable incompleteness of sensor-based models due to limited sensor scope, sensor noise, and discretization effects. Heretofore, most robot systems designers have either ignored this uncertainty completely, or have used various ad-hoc or approximate methods of representing or accounting for it. In either case, data is accepted passively with no consideration as to how uncertainty may be reduced or avoided. This often forces decisions about action to be made conservatively in order to account for possible sensor errors or omissions. In the long run, this policy seriously affects the performance of the system.

Substantial gains result from explicitly taking action to reduce uncertainty by using *active sensing*[2]. This removes the restriction of a predetermined data set and replaces it with the more flexible approach of actively seeking and using several scenes or samples. Of course, this adds the complication of modeling sensors [14] and constructing procedures to control sensors.

The specific problem of sensor control has not received much attention, but it is analogous to art of *experimental design* [3, Chap. 7] and [16]. The science of experiment design focuses on maximizing the information obtained from an experimental program under resource constraints. In our case, the sensor is a measurement system with control parameters. We will assume we are given a set within which the parameter lies, and we seek to improve this estimate. Our objective is to find a plan of control will give the "best" post-experimental results within the constraints of time and processor resources.

Our problem setting includes the following important attributes: First, the systems we consider are nonlinear in both state and control. Second, our measurement noise depends on the control of the measurement system. Third, our control criterion is a direct function of the information returned by the estimation procedure. Finally, and perhaps most importantly, our information is limited by sensor scope. Our information may vary widely and discontinuously based on what view is presented to the sensory system.

In this paper, we will focus on statistically based estimation and control techniques for dealing with sensor noise in nonlinear measurement systems. We will first discuss what constitutes a suitable payoff function for sensor control. Then we will evaluate several estimation techniques. Finally, we will discuss what sampling strategies are appropriate and what additional complication control introduces.

## 2 A Sensor Control Formulation

Recovering information from a noisy sensor is a problem in statistical estimation. We shall argue that control should be based on the expected or posterior risk of the estimation procedure. This in turn motivates the requirement that estimation procedures not only estimate well, but have predictable performance.

*Acknowledgements: This work was supported in part by NSF/DCR 8410771, NSF DMC-8411879 and DMC-12838, US Air Force F49620-85-K-0018, US Air Force F33615-83-C-3000, US Air Force F33615-86-C-3610, DARPA/ONR, ARMY/DAAG-29-84-K-0061, NSF-CER DCR82-19196 A02, NIH NS-10939-11 as part of the Cerebrovascular Research Center, by DEC Corp., and LORD Corp.

## 2.1 Formalizing the Control Problem

### 2.1.1 Estimation

A sensor takes observations of the environment as described by some transfer function contaminated by noise. Both the transfer function and the observation noise can be influenced by control parameters. Thus, we can formalize a controllable measurement device subject to additive noise as a mathematical system of the following general form:

$$z_i = H(u_i, p) + V(u_i, p) \quad (1)$$

where $H$ is $k$-dimensional, $u_i$ is an $m$-dimensional control vector from a set $\mathcal{U}$, and $p \in \mathcal{P}$ is the $s$-dimensional quantity we are attempting to estimate. We observe $z_i$, a function of both $u_i$ and $p$ contaminated by additive noise $V(\cdot, \cdot)$ of dimension $k$. In general, the distribution of $V$ will be a function of our control parameter and the parameter of interest. Our problem is to optimize, by choice of some sequence $\underline{u} = [u_1, u_2, \ldots, u_n] \in \mathcal{U}^n$, the performance of an estimation procedure $\delta_n(\cdot)$ estimating $p$ from $\underline{z} = [z_1, z_2, \ldots, z_n]$.

In order to choose a particular estimation procedure, we must pick a criterion or *loss* by which to judge the merit of a decision rule. A reasonable and commonly assumed loss criterion which often leads to simple, iterative estimation rules is the *mean square error* loss. The optimal estimation procedure is that which, given a prior $\pi(\cdot)$ on $p$ and a conditional distribution on the observations, $f(\cdot|p)$, minimizes the quantity

$$E \|\delta(z) - p\|^2 \quad (2)$$

If $p$ is bounded, we know that the $\delta$ which minimizes this expression is

$$\delta(z) = E[p|z] \quad (3)$$

This is the Bayes solution to our estimation problem [3]. We shall henceforth assume that some estimation procedure, $\delta$, implementing or approximating the solution to Equation 3 is given *a priori*.

### 2.1.2 Control

A control sequence is to be evaluated relative to its expected utility. This utility can be thought of in two parts: the performance of the estimation procedure for that choice of strategy, and the cost of implementing that strategy. This can be expressed as

$$l(p, \hat{p}, n, \underline{u}) = l^d(p, \hat{p}) + c(n, \underline{u}) \quad (4)$$

$l^d$ represents the loss attributed to the estimation procedure $\delta$ and $c$ represents the cost of taking $n$ samples via the control strategy $\underline{u}$.

The choice of actual functional forms for $c$ and $l^d$ reflects the desired behavior of the system, so before choosing $c$ and $l^d$ we must consider what role sensors are to play. In earlier reports [9,10,11,12], we argue that sensors should be independent devices with dedicated computational resources and substantial local intelligence. We present a general organizational structure for robot systems in which intelligent sensors are organized for specific tasks by higher level agents while functioning with as much autonomy as possible. This organization is coordinated by managers who send agents task requests, and integrate sensor results into a world model.

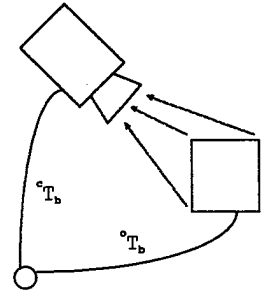

Figure 1: The sensor positioning problem.

The format of a sensor request consists of three elements. The quality of the information requested is expressed by a tolerance, $\epsilon$. A better estimate can always be gained by using more samples, but at a higher cost. A priority, $w$, encodes a time/accuracy tradeoff for the requested information. Finally, a query will contain a hard time constraint, $t$, which should not be violated. Once a sensor performs a sensing task, the results returned should indicate the degree to which the set task was found to be achievable. Since we have adopted probability distributions as a means of characterizing the system's information about $p$, it is natural that the sensor should return a probability indicating the quality of the final estimate. In summary, a sensor request consists of a triple $\langle \epsilon, t, w \rangle$, and sensor results take the form of a pair $\langle \hat{p}, P(|\hat{p} - p| < \epsilon|\underline{z}) \rangle$.

Our choice of an uncertainty region, $\epsilon$, suggest that an appropriate evaluation criterion for the performance of an estimation procedure $\delta_n$ in the context of control is the *0-1 loss*.[1]

$$l^d(p, \delta_n(\underline{z})) = \begin{cases} 0, & \text{if } |\delta_n(\underline{z}) - p| < \epsilon; \\ w, & \text{otherwise} \end{cases} \quad (5)$$

By taking expectations of Equation 4, we can compute the *Bayes decision risk* of an estimation procedure $\delta_n$ as

$$r(n, \underline{u}, \delta) = E\left[l^d(p, \delta_n(z)) + c(n, \underline{u})\right] = r^d(\pi, \delta_n) + c(n, \underline{u}) \quad (6)$$

It is easy to show that $r^d(\pi, \delta_n) = P(|\delta_n(\underline{z}) - p| > \epsilon)$. In other words, the Bayes decision risk associated with the procedure $\delta_n(\cdot)$ is the probability of the estimated value falling outside the specified tolerance.

### 2.1.3 An Example Problem

In order to illustrate the above concepts, consider estimating the position of an object using a stereo camera as illustrated in Figure 1. For ease of exposition, we will only consider the 2-D case using orthogonal projection. In this case, the sensor can by modeled by the combined transformation from object coordinates to camera coordinates.

---

[1] It is important to realize that $\delta_n$ may have been *derived* from very different criteria, *i.e.*, the mean square error criterion. Now, *given* $\delta_n$, we are evaluating it relative to the 0-w loss.

$$^cT_o(p,u) = {^cT_b(u)}{^oT_b}^{-1}(p) \qquad (7)$$

Let $^of_i = [x_i, y_i, 0, 1]^T$ denote a homogeneous feature position in object coordinates, and let $^cz_i = [x_i, y_i, 0, 1]^T$ represent an observation of $f_i$ in camera coordinates. The vector $p$ we are estimating is $p = [x_o, y_o, \alpha_o]^T$ of an object. Our control vector, $u = [x_c, y_c, \alpha_c]$, corresponds to the choice of camera position. The control problem is essentially a choice of how to move the camera about an object (which lies within some region of uncertainty) so as to achieve the best final estimate of position.

## 2.2 Review of Decision Theory

If we fix $\underline{u}$ in Equation 6, then we can consider the problem of finding the n which minimizes $r(n, \underline{u}, \delta)$. A procedure where $n$ samples are taken without looking at any intermediate results is referred to as a *batch procedure*. Notice, the optimal control sequence is then that which minimizes the batch size for a given posterior loss. On the other hand could also solve this problem by evaluating the risk conditioned on the observed data, and deriving a *stopping rule* which determines when enough data has been taken. This is a more efficient procedure, but sequential problems tend to be more difficult than batch problems. Berger [3] provides a good overview Bayesian sequential and batch decision problems. Results are analytically derivable in some special cases, but there is no general theory for problems of the scope considered in our setting.

There are many unanswered questions as to how estimation and control interact. It is well known in control theory that, in the case of a linear dynamic system with a gaussian prior on the initial state and observed via linear measurement equations with additive gaussian noise, estimation and control separate into independent problems [4]; and that the *Kalman Filter* is the *optimal* estimation procedure. If either the observation noise or prior are not gaussian, the Kalman filter is still the optimal *linear* estimation procedure. In the case where the system state is static (*i.e.*, a parameter), the Kalman filter becomes a iterative implementation of Bayes estimation, and the solution to Equation 3 can be written as

$$\begin{aligned}
\hat{p}_{i+1} = E[p|\underline{z}_i] &= (I - K_i H_i)\hat{p}_i + K_i z_i, & (8) \\
\Lambda_{i+1} &= (I - K_i H_i)\Lambda_i & (9) \\
K_i &= \Lambda_i H_i^T (H_i \Lambda_i H_i^T + \Lambda_v)^{-1}
\end{aligned}$$

Sequential and batch rules for sequential Bayes estimation are easily derived based on the magnitude of the error terms [3]. Müller and Weber [17] consider the problem of finding the measurement system design maximizing a suitable norm of the observability or controllability of a system linear in both state and control. The norms they discuss are the trace, determinant, an maximum eigenvalue of the observability matrix. Mehra [15] combines and extends these results to include time varying systems and randomized designs.

Meier [13] deals with specializations of Equation 1 in which $V(\cdot)$ is constant zero-mean gaussian noise, and the system is linear in $p$ as in

$$z = H(u)p + V$$

In this case, the best measurement control to achieve to minimize the control criterion of the driving system can be derived. It is shown that the optimal control is open-loop and the solution is obtained by a dynamic programming algorithm.

There are not many general solutions for nonlinear measurement systems, though there are a number of approximation techniques [8]. One approximation is to linearize a given nonlinear measurement system about a nominal trajectory and apply linear estimation techniques. In the case of the Kalman filter, the resulting quasi-linear procedure is referred to as the *Extended Kalman filter* (EKF). Anther method of nonlinear estimation which does not rely on linearization is *Stochastic Approximation* [18]. This is an asymptotic technique akin to Newton's method adapted to work in the presence of noise. The technique can be written as a recurrence of the form

$$p_{n+1} = p_n + a_n(z_n - h(p_n))$$

The choice of gain sequences $a_1, a_2, \ldots$ is crucial for the convergence of the technique. Asymptotically optimal control results exist for stochastic approximation, but the small-sample behavior of the estimation/control method are not known. Since no priors are assumed it is difficult to derive a good measure of the rate of convergence.

In our case, it is reasonable to assume some prior knowledge about both the parameter of interest and statistical properties of sensor noise. Hence we will seek a Bayesian solution to the problem. This naturally leads us to consider the Kalman filter.

## 3 The Extended Kalman Filter

Kalman filters and Extended Kalman filters have a long and well documented history of use in navigation problems very similar to our sample problem [20]. In the field of robotics, Durrant-Whyte [6] has used a version of this technique to solve the problem of updating location estimates from observations. Smith et. al. [19] have worked this out for a mobile robot estimating position. Ayache and Faugeras [1] have looked at several problems in stereo ranging by building an EKF from a general constraint equation.

The EKF is derived in the following fashion. Given a nonlinear, measurement system such as Equation 1, we can approximate with the first order terms of a Taylor series centered about a prior estimate for $p$, $\hat{p}$, and apply a Kalman filter to the resulting quasi-linear system. The resulting system equations (found in many references, *e.g.* [8]) is $y = Mp + V$ where we compute $M$ and $y$ from

$$M = \frac{\partial H(\hat{p}, u)}{\partial p}$$
$$y = z - H(u, \hat{p}) + \frac{\partial H(u, \hat{p})}{\partial p}^T \hat{p} \quad (10)$$

The $0 - w$ risk of the EKF can be computed by evaluating the probability mass that lies outside the tolerance set based on the error term, $\Lambda_i$. This risk may be used to select a batch size, considered sequentially in a stopping rule, or be used to pick control points. In [11], we implemented the EKF for our sample problem and computed the expected information when observing a simple polygonal object from all possible sensor views. However, it is important to note that Equation 10 *depends* on the value we are estimating — *i.e.* it is a random variable itself. Slight errors in the linearization can substantially affect how well the error terms represent the true error of the filter.

The possible errors in the linearization can be thought of as uncertainty about the correct parameterization of the filter. Several papers in the literature analyze the sensitivity of the Kalman filter with regard to uncertainty in filter parameters [7]. When a dynamic system is involved, two types of deficiencies are recognized: *divergence* and *true divergence*. The former is a local deviation of the filter to a steady state where errors remain bounded, but the filter does not converge to the correct values. The latter is the case where error terms become unbounded. Most small errors can be overcome by adding extra observation noise, though this slows the convergence of the filter. In our case, we can compute the worst bias introduced by the error in linearization and feed this back as an error term. However, for any substantial uncertainty in the original parameter, the filter converges extremely slowly.

In the static cases such as the one we are considering, true divergence will never happen. Still, if the previous estimate is poor, then we can expect that the linearization, and hence the information maps, diverge substantially from the true values. In fact, in real practice it can turn out that even the static filter *does not even converge in the limit*! This is because the EKF, like stochastic approximation, is a differential correction technique. If the correction factor goes to zero too quickly, the true solution may be excluded. This can be seen by substituting Equation 10 into Equation 8. For a scalar system, this yields:

$$E[p|z] \approx (1 - k h')\hat{p} + k (z - h + h'\hat{p})$$
$$= \hat{p} + k(z - h) = \hat{p} + \frac{\sigma_p^2 h'}{h'^2 \sigma_p^2 + \sigma_v^2}(z - h)$$

where $h' = h'(u, \hat{p})$ and $h = h(u, \hat{p})$. Depending on the values of $h'$, $\sigma_p^2$ and $\sigma_v^2$, $k$ may or may not be an appropriate gain to choose for convergence. Discovering when $k$ is appropriate will delimit when the computed variance/covariance matrices are good approximations.

| Bound | Est. Err. | Obs. Err. | % Error |
|---|---|---|---|
| ± 0.1 | 0.0033 | 0.0031 | 7% |
| ± 0.2 | 0.0134 | 0.0142 | -6% |
| ± 0.3 | 0.0306 | 0.0417 | -27% |
| ± 0.4 | 0.0552 | 0.0708 | -22% |
| ± 0.5 | 0.0882 | 0.1772 | -50% |
| ± 0.6 | 0.1298 | 0.2981 | -56% |
| ± 0.7 | 0.1839 | 0.5070 | -64% |

Table 1: A comparison of the EKF term for $\alpha_o$ with observed estimation error of $\alpha_o$ in a simulation.

Table 1 illustrates the problem.. This data was generated by running the EKF for the example problem. Values for $x_o$ and $y_o$ were held fixed at mean values, and the interval of uncertainty of $\alpha_o$ was varied from $\pm .1$ to about $\pm \pi/4$. It can be seen that as the interval of uncertainty widens, the error terms begin to drastically under-represent the error in estimation. For large enough intervals, the filter often fails to even converge for all practical purposes.

We will analyze the sensitivity of the linear filter via game-theoretic techniques. Consider a scalar version of Equation 1. For any given point, $p$, and linearization point, $\hat{p}$, the linear form

$$z(\eta) = h(u, \hat{p}) + \frac{\partial h(u, \eta)}{\partial p}(p - \hat{p}) + v$$

is *exact* for some $\eta$ between $p$ and $\hat{p}$. Instead of building a linear procedure based on assuming $z(\hat{p})$, we can build the procedure based on the worst case value of $\eta$. This can be stated as

$$\min_a E[\max_\eta (a \ z(\eta) - p)^2]$$

At this point, $\eta$ is restricted to fall between $p$ and the linearization point $\hat{p}$. We can rewrite this in more standard form by moving the maximization outside the expectations and replacing $\eta$ by a function $\eta(\cdot)$ representing the constraint on $\hat{p}$:

$$\min_a \max_{\eta(\cdot)} E[(a \ z(\eta(p)) - p)^2] \quad a \in \mathcal{A}$$

where we now maximize over a set of functions $\eta \in \mathcal{E}_{\hat{p}}$. To remain consistent, we must restrict $\mathcal{E}_{\hat{p}}$ to contain only those functions which map a point $p$ into the interval between $p$ and $\hat{p}$. Furthermore, let $g(\cdot) = z(\eta(\cdot))$ and let $\mathcal{G}$ denote the set of all such $g^2$ so that Equation 3 becomes

$$\min_a \max_g E[(a \ g(p) - p)^2] \quad g \in \mathcal{G}, \quad a \in \mathcal{A} \quad (11)$$

We now ask if there is a *single* worst case $g$ and best $a$ as the solution to this problem (in the language of game theory, is there a nonrandomized saddle-point solution). To answer, we appeal to a theorem due to Blackwell and Gershick:

---
[2]We will suppress indicating $\hat{p}$ from now on.

**Theorem 3.1** Let the triple $(K, \mathcal{X}, \mathcal{Y})$ denote a two-person zero-sum game where

- $X$ is a compact convex subset of $E^m$
- $Y$ is a compact subset of $E^n$
- $K(x, y)$ is a real-valued mapping defined on the Cartesian product of $X$ and $Y$
- for each $y \in \mathcal{Y}$, $K(x, y)$ is convex in $x \in \mathcal{X}$
- for each $y \in \mathcal{Y}$, $K(x, y)$ is continuous in $x \in \mathcal{X}$
- for each $x \in \mathcal{X}$, $K(x, y)$ is continuous in $y \in \mathcal{Y}$

Then the game $(K, \mathcal{X}, \mathcal{Y})$ has a saddle-point solution $(x^*, \lambda^*)$

$$\min_{x \in \mathcal{X}} \max_{y \in \mathcal{Y}} E_\lambda(K(x,y)) = \max_{y \in \mathcal{Y}} \min_{x \in \mathcal{X}} E_\lambda(K(x,y))$$
$$= E_{\lambda^*}(K(x^*, y))$$
$$= \int_{\mathcal{Y}} K(x^*, y) \, d\lambda^*$$

where $x^* \in \mathcal{X}$, $\lambda^* \in \mathcal{M}_\mathcal{Y}$, $\mathcal{M}_\mathcal{Y}$ denotes the class of all probability measures on $\mathcal{Y}$, and $\lambda^*$ is a discrete probability distribution which assigns mass to at most $m+1$ points in $\mathcal{Y}$.

The proof of Theorem 3.1 can be found in [5]. We note that, if $\mathcal{Y}$ is convex and $K(x, y)$ is strictly convex in $y$, then the support of $\lambda^*$ must lie on the boundary of $\mathcal{Y}$.

It is easy to show that the set of functions $\mathcal{E}$ is convex as we have defined it. As it turns out, $\mathcal{G}$ is also convex for any continuous function $z(\cdot)$. However, the statement of the theorem as we have it here applies to Euclidean spaces, not function spaces.[3] In order to apply the theorem, we make heuristic use of a discrete (vector) approximation for the functions in $\mathcal{G}$. Let $\Gamma = [\gamma_1, \gamma_2, \ldots, \gamma_n]^T$ be a sequence of approximation points falling in the set $\mathcal{P}$. Let $\mathcal{G}' = \{[g(\gamma_1), g(\gamma_2), \ldots, g(\gamma_n)]^T | g \in \mathcal{G}\}$. Then, $\mathcal{G}'$ is a compact, convex subset of $\mathcal{E}^n$ as required for the theorem.

If we define $K(a, g) = E[(a\ g(p) - p)^2]$ where the prior on $p$ now assigns mass only to points in $\Gamma$, then the maximizer is essentially choosing an $n$-vector in $\mathcal{G}'$. $K$ is easily seen to be a continuous function which is convex in both arguments. Hence, the game $(K, \mathcal{A}, \mathcal{G}')$ satisfies the conditions of Theorem 3.1. Moreover, since $K$ is convex in $g'$, the saddle-point solution comes from the boundary of $\mathcal{G}'$ or $\mathcal{M}_{\mathcal{G}'}$ as appropriate.

We stated that we wanted a *single* worst case $g$ – that is, we require that $g^* \in \mathcal{G}$. A qualitative, analysis reveals the choice of a nonrandom or random solution is affected by the size of $\mathcal{P}$, and the signal/noise ratio. If the uncertainty set is large, or the signal/noise ratio is good, the optimal maximizer strategy is likely to be a randomized strategy. The need for a randomized solution complicates the design of the optimal EKF and gives us cause to wonder if the EKF based on linearizing about the (single) prior estimate is a reasonable approximation.

---
[3] It can be strengthened to include the function space we are interested in. However, this result requires substantially more space and mathematical machinery, and does not lend any more insight than the approach we are taking here.

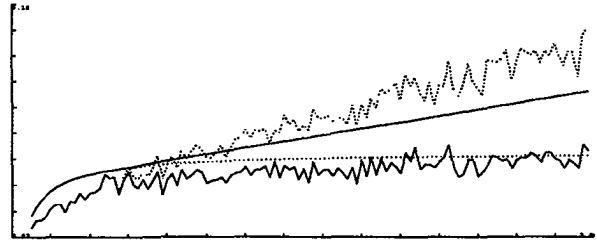

Figure 2: A comparison of the error terms of the game-theoretic filter (solid lines) to the error terms of a lower bound filter (dotted lines). The smooth curve represents the calculated error, while the jagged lines are the error observed in a simulation.

For a specific example, consider a simple linear system, $z = h\theta + v$ where $h$ comes from some set $\mathcal{H}$. In this case, we can apply the saddle-point theorem directly and observe that the saddle-point solution is a linear combination of the boundary points of $\mathcal{H}$. In this case it is straightforward to evaluate $K(x, y)$ and discover whether or not the optimal maximizer strategy for a given set of parameters is random and, if so, what the strategy is. For example, if $\mathcal{H} = [4, 5]$ and $\sigma_v^2 = 1$, then it is possible to show that optimal filter is randomized for $\sigma_\theta^2 > .6$. Figure 2 compares the behavior of the game-theoretic filter with a filter using the lower bound of the $\mathcal{G}$ set by varying over $\sigma_\theta^2$. Both filters agree until the game filter (solid lines) uses a random strategy, at which point the error terms of the lower bound (dotted lines) filter begin to underestimate the filter error. The game filter error terms continue to behave well.

The generalization of this result to the multivariate case, and its utility in predicting the performance of the EKF will be presented in an extended version of this paper.

## 4 Approximate Bayesian Estimation

Given a series of observations, $\underline{z} = [z_1, z_2, \ldots, z_n]$, from a system with additive noise distributed as $f_v(\cdot)$ and prior $f_p(\cdot)$, the Bayesian solution for any estimation problem is given by Equation 3. In general this solution can be only be computed numerically. Moreover, this procedure cannot be realized sequentially.

If $f_v(\cdot) \sim N(\mu, \Lambda)$ where the individual components are independent (i.e., $\Lambda$ is diagonal), then it is well known that the sample mean, $\overline{z}$, is again gaussian with covariance $\Lambda/n$. In this case, we can rewrite Equation 3 as

$$\delta_n(\underline{z}) = \delta(\overline{z}) = \int_\mathcal{P} p\ f(p|\overline{z})\ dp = \frac{\int_\mathcal{P} p\ f(\overline{z}|p)\ \pi(p)\ dp}{\int_\mathcal{P} f(\overline{z}|p)\ \pi(p)\ dp} \quad (12)$$

| Bound | Obs. Err. | % of EKF |
|---|---|---|
| ± 0.1 | 0.0030 | 95% |
| ± 0.2 | 0.0096 | 67% |
| ± 0.3 | 0.0238 | 57% |
| ± 0.4 | 0.0428 | 60% |
| ± 0.5 | 0.0375 | 21% |
| ± 0.6 | 0.0824 | 28% |
| ± 0.7 | 0.1076 | 21% |

Table 2: The observed mean square error in $\alpha_o$ of the Bayes approximation and its ratio to the observed mean square error of the EKF.

If the prior, $\pi(\cdot)$, is chosen to have a continuous cumulative density function, then numerical integration appears to be the only solution to Equation 12. However, if we choose $\pi(\cdot)$ to have a discrete c.d.f, i.e., $\pi(\cdot)$ assigns mass to a finite number of atoms, then we achieve a substantial decrease in the complexity of computing Equation 12. Since we only need to consider a finite set of atoms, the integrals become weighted sums.

$$\delta^{lf}(\overline{z}) = \frac{\sum_{j_1} \sum_{j_2} \cdots \sum_{j_n} p \, f(\overline{z}|p^T) \, \pi(p)}{\sum_{j_1} \sum_{j_2} \cdots \sum_{j_n} f(\overline{z}|p^T) \, \pi(p)} \quad (13)$$

where $p = [p_{j_1}, p_{j_2}, \ldots, p_{j_n}]^T$

From a computational standpoint, it is important to note that any constants of integration divide out in the fraction. Moreover, if the prior on $p$ is uniform, then the weighting constants are all the same, so no multiplies are required. Also, note that, by considering the grid elements, the condition distribution values, and the prior values as vectors, the nested sums can be computed by a simple dot project operation.

### 4.1 Finite Prior Bayes Estimation

We have implemented both scalar and multivariate versions of this estimation procedure. Table 2 shows the observed mean-squared-error of the EKF vs. the observed error of the Bayes approximation with a 5-point support for the same situation as described in Table 1. The first column lists the observed error of the Bayes approximation, while the second column lists the ratio of Bayes observed error to EKF observed error. We see that the Bayes procedure yields a substantial performance improvement over the EKF for all intervals investigated.

This is in rough correspondence with the results of the previous section. If we were to compute the interval width at which the solution to Equation 11 is randomized, we would find that it occurs at very small intervals.[4]

We can compute the risk of $\delta$ conditioned on the observed data by employing Equation 13 with $(p - \delta(\overline{z}))^2$

---

[4] We can obtain an upper bound by neglecting the restriction that $\eta$ falls between $p$ and $\tilde{p}$ and generalizing the results to the multivariate case. If the solution to this problem requires randomization, the solution to Equation 11 does also.

| Iter | Bound | Computed | Real | % Diff |
|---|---|---|---|---|
| 3 | 0.1 | 0.0019 | 0.0017 | 13% |
| 3 | 0.2 | 0.0029 | 0.0027 | 7% |
| 3 | 0.3 | 0.0034 | 0.0030 | 14% |
| 3 | 0.4 | 0.0035 | 0.0039 | -10% |
| 5 | 0.1 | 0.0014 | 0.0015 | -4% |
| 5 | 0.2 | 0.0018 | 0.0019 | -4% |
| 5 | 0.3 | 0.0021 | 0.0024 | -13% |
| 5 | 0.4 | 0.0024 | 0.0035 | -31% |
| 7 | 0.1 | 0.0012 | 0.0011 | 11% |
| 7 | 0.2 | 0.0014 | 0.0015 | -11% |
| 7 | 0.3 | 0.0015 | 0.0021 | -27% |
| 7 | 0.4 | 0.0013 | 0.0034 | -61% |

Table 3: A comparison of the computed posterior squared error of the Bayes Approximation to the observed error in a simulation.

substituted for $p$. Table 3 shows the results of a scalar simulation for $z = 10 \sin(p) + v$ where $v$ is a gaussian random variable with $\sigma_v^2 = 1$, and $p$ has a uniform prior within ± Bound. The percentages in the righthand column are the difference between the estimated error and the observed simulation error divided by the simulation error. Positive percentages indicate the computed error exceeded the observed error, and negative percentages indicate the opposite. Notice that the performance of the error estimator degrades as a function of both iterations and interval width. This also suggests that the performance of the estimator itself is degrading.

The reason for this degraded performance can be best illustrated by Figure 3. This is a plot of the expected value of our Bayes approximation for this system with a uniform prior over $\mathcal{P} = [-\pi/2, \pi/2]$. The approximation is close to the actual Bayes procedure until the signal to noise ratio becomes good, at which point it begins to exhibit stepwise behavior. This is to be expected since, in the limit, the estimation procedure becomes a *quantizer*. Since variance of the sample mean is inversely related to sample size, the procedure exhibits quantization for large sample sizes. Similarly, as the interval gets larger, a fixed number of prior points become a poorer approximation of the true prior.

Obviously, if the procedure begins to quantize, one solution is to use a larger number of prior points. However, the complexity of this method is $O(r^m)$ where $r$ is the number of prior points and $m$ is the dimensionality of $p$. For high dimensional problems, the computation for more than 2 or 3 points is overwhelming.

### 4.2 Iterative Finite Prior Bayes Approximation

The quantization effects of the Bayes approximation can be used to construct an iterative approximation by the following method: Take samples until the method quantizes. Divide the interval about the selected prior and continue to sample until this smaller interval quantizes. This can be continued indefinitely, though with increasing risk for fixed sample size.

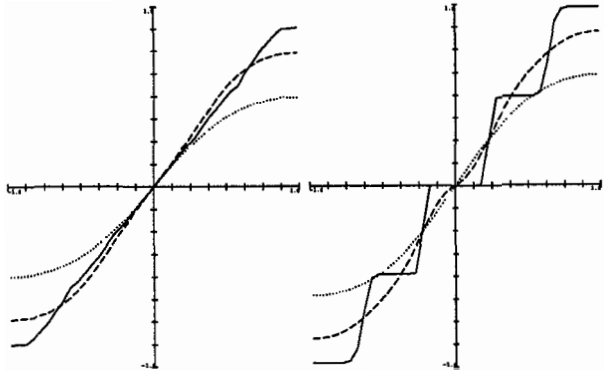

Figure 3: A numerical evaluation of the Bayes estimation procedure (left) vs. the Bayes approximation procedure (right) for $f(x) = 10\sin(x)$. The dotted line represents the results for $\sigma_{\bar{z}}^2 = 30$, the dashed line represents $\sigma^2 = 3$, and the solid line represents $\sigma^2 = .1$.

More precisely, let the interval notation $[a, b]$ denote a generalized interval, i.e $[a, b] = \{p | a_i \leq p_i \leq b_i, i = 1, \ldots, n\}$. Define the *left neighbor* of a scalar prior point $p_j$ as $p_{j-1}$ if $j > 1$ and $p_j$ if $j = 1$. The right neighbor of a point is the obvious dual. The left and right neighbors of a vector are the vectors of left and right neighbors of its elements. Then, the *neighborhood* of a prior $p$ is the interval $[a, b]$ where $a$ is the left neighbor of $p$ and $b$ is right neighbor. The neighborhood of a point which is not a prior point is the neighborhood of the closest prior point. Finally, $\mathcal{P}_{a,b}$ will denote the set of grid points for the interval $[a, b]$. Then the iterative procedure can be written as:

```
Finite_Bayes(z, [a_i, b_i]) :=
    compute P_{a_i,b_i} as the new prior
    p_i = δ^{lf}(z)
    If b_i - a_i < ε then
        ⟨p_i, P(p_i ∈ [a, b])⟩
    else
        set a_{i+1} to the left neighbor of p_i
        set b_{i+1} to the right neighbor of p_i
        Finite_Bayes(z, [a_{i+1}, b_{i+1}])
```

We implement this algorithm using 5pt priors – thus reducing the size of uncertainty set by a factor of two at each step. If $c_1$ is the cost of running the estimation procedure, $l$ is the length of the original interval, and $\epsilon$ then is the requested tolerance, then the cost of computing an estimate with this approximation procedure is

$$c_1 \lceil \log_2 l/\epsilon \rceil + c_2(\underline{z})$$

where $c_2(\underline{z})$ represents the sampling cost for the observations $\underline{z}$.

If $p$ is estimated to lie in interval $\mathcal{I}$, then the conditional probability of error in the scalar case, $P(|\hat{p} - p| > \epsilon|\bar{z})$, is given by

$$\int_{\mathcal{P}-\mathcal{I}} f(p|\bar{z})\, \pi(p)\, dp \qquad (14)$$

In the general nonlinear case, this integral must be computed numerically, though certain heuristics can speed it up.

### 4.3 Batch Rules and Control

Linear-gaussian systems enjoy the property that batch rules and control laws are generally open-loop (recall the work of Meier). In the case of non-linear or non-Gaussian systems, batch size or control will often depend on the parameter under observation. Hence, optimality is much more difficult to attain, especially in a time-constrained situation.

Note that the error terms of all of the procedures mentioned above depend on the parameter under observation. Since a batch procedure uses only prior information, a sequential experimental design using intermediate information about the parameter seems more appropriate. However, the risk of the EKF is subject to error, and the risk of the approximation procedures is too expensive to compute at each sample. The optimal design is some combination of a batch procedure and a sequential procedure.

For the iterative Bayes approximation, the natural solution is to compute the batch size to the next quantization for a given risk level. When data is taken during estimation, the estimation risk of the sequential approximation is the product of the risk for each estimate. That is, if our target risk is $1 - r$, and the number of iterations required is $i$, we will choose the target risk of each stage to be $1 - \sqrt[i]{r}$. [5] Given a current estimate $p$, we choose the batch size, conditioned on $p$, which will force quantization. This leads to a series of batches, the size of each conditioned on the latest value of $p$.

For control, we use batch size as the evaluation criteria. We define the cost of a control point as the time to take the required batch of samples plus the travel time to that point. The optimal point is that which requires the least amount of time.

## 5 Conclusions and Discussion

We have presented a mathematical formulation of the sensor control problem. The solution to this problem requires an accurate and predictable estimation technique. We have two techniques, the extended Kalman filter, and a finite Bayes approximation. We have found via a game-theoretical analysis that the EKF is very sensitive to the size of the uncertainty interval. This correlates with the observed behavior of an EKF running on a sample problem. The Bayes approximation does not appear to suffer from these deficiencies. Of these two, the latter appears to be a more robust and predictable estimation technique.

---
[5] Clearly, the estimate will exhibit better localization properties if estimation is carried out after all of the data is taken. Conversely, it is better from a control perspective to estimate as soon as possible since the information gained may allow better control. The best tradeoff seems to estimate at uniform intervals.

There are several questions still to be addressed in evaluating sensor estimation techniques. For instance, what is the time/accuracy tradeoff between an EKF and the finite Bayes technique? Is the Bayes approximation always better, worse or mixed for the same set of samples? What sorts of parallelism can be exploited? From a theoretical perspective, we need to further evaluate our game-theoretic studies and extend them to the general case.

We are currently working on the control problem for sequential and batch estimation scenarios. To solve the control problem, we will have to face new issues dealing with the evaluation of experimental risk, and choosing the next view. We also plan to extend these procedures for multi-sensor estimation and control. Equation 1 can be easily generalized for $n$ sensors all observing and correlating information. However, multi-sensor control requires the arbitration of communication and negotiation.

Finally, we wish to consider the case where the prior knowledge is not a parametric description as we have supposed, but a structural one. In this case, resolving uncertainty involves reasoning about structural and environmental constraints. The higher level will use the low level procedures we have developed here to implement search strategies.